\documentclass[journal=apchd5,manuscript=article]{achemso}

\usepackage[version=3]{mhchem} 
\usepackage{graphicx}
\usepackage{color}
\usepackage{amsxtra}
\usepackage{amstext}
\usepackage{latexsym}
\usepackage[squaren]{SIunits}
\usepackage{natbib}
\usepackage{sidecap}
\usepackage{floatflt}
\usepackage{ulem}

\author{Samuel M.H. Luk}
\affiliation[UoA]{Department of Physics, University of Arizona, Tucson, AZ 85721, USA}
\author{Hadrien Vergnet}
\affiliation[LPENS]{Laboratoire de Physique de l'Ecole Normale Sup\'{e}rieure, ENS, Universit\'{e} PSL, CNRS, Sorbonne Universit\'e,
Universit\'{e} Paris-Diderot, Sorbonne Paris Cit\'{e}, 24 rue Lhomond 75005 Paris, France}
\author{Ombline Lafont}
\affiliation[LPENS]{Laboratoire de Physique de l'Ecole Normale Sup\'{e}rieure, ENS, Universit\'{e} PSL, CNRS, Sorbonne Universit\'e,
Universit\'{e} Paris-Diderot, Sorbonne Paris Cit\'{e}, 24 rue Lhomond 75005 Paris, France}
\author{Przemyslaw Lewandowski}
\affiliation[Paderborn]{Physics Department and Center for Optoelectronics and Photonics Paderborn (CeOPP), Universit\"at Paderborn, Warburger Strasse 100, 33098 Paderborn, Germany}
\author{Nai H. Kwong}
\affiliation[Paderborn]{Wyant College of Optical Sciences, University of Arizona, Tucson, AZ 85721, USA}
\author{Elisabeth Galopin}
\affiliation[C2N]{Centre de Nanosciences et de Nanotechnologies (C2N), CNRS, Universit\'e Paris Sud, Universit\'e Paris-Saclay, 91120 Palaiseau, France}
\author{Aristide Lemaitre}
\affiliation[C2N]{Centre de Nanosciences et de Nanotechnologies (C2N), CNRS, Universit\'e Paris Sud, Universit\'e Paris-Saclay, 91120 Palaiseau, France}
\author{Philippe Roussignol}
\affiliation[LPENS]{Laboratoire de Physique de l'Ecole Normale Sup\'{e}rieure, ENS, Universit\'{e} PSL, CNRS, Sorbonne Universit\'e,
Universit\'{e} Paris-Diderot, Sorbonne Paris Cit\'{e}, 24 rue Lhomond 75005 Paris, France}
\author{Jérôme Tignon}
\affiliation[LPENS]{Laboratoire de Physique de l'Ecole Normale Sup\'{e}rieure, ENS, Universit\'{e} PSL, CNRS, Sorbonne Universit\'e,
Universit\'{e} Paris-Diderot, Sorbonne Paris Cit\'{e}, 24 rue Lhomond 75005 Paris, France}
\author{Stefan Schumacher}
\affiliation[Paderborn]{Physics Department and Center for Optoelectronics and Photonics Paderborn (CeOPP), Universit\"at Paderborn, Warburger Strasse 100, 33098 Paderborn, Germany}
\alsoaffiliation[Paderborn]{Wyant College of Optical Sciences, University of Arizona, Tucson, AZ 85721, USA}
\author{Rolf Binder}
\affiliation[UoA]{Department of Physics, University of Arizona, Tucson, AZ 85721, USA}
\alsoaffiliation[Paderborn]{Wyant College of Optical Sciences, University of Arizona, Tucson, AZ 85721, USA}
\author{Emmanuel Baudin}
\affiliation[LPENS]{Laboratoire de Physique de l'Ecole Normale Sup\'{e}rieure, ENS, Universit\'{e} PSL, CNRS, Sorbonne Universit\'e,
Universit\'{e} Paris-Diderot, Sorbonne Paris Cit\'{e}, 24 rue Lhomond 75005 Paris, France}
\email{baudin@phys.ens.fr}

\title[All-optical beam steering using the polariton lighthouse effect]
  {All-optical beam steering using the polariton lighthouse effect}


\begin{document}
\begin{abstract}
  We demonstrate theoretically and experimentally that a specifically designed microcavity driven in the optical parametric oscillation regime exhibits lighthouse-like emission, i.e., an emission focused around a single direction. Remarkably, the emission direction of this micro-lighthouse is continuously controlled by the linear polarization of the incident laser, and angular beam steering over \unit{360}{\degree} is demonstrated. 
Theoretically, this unprecedented effect arises from the interplay between the nonlinear optical response of microcavity exciton-polaritons, the difference in the subcavities forming the microcavity, and the rotational invariance of the device.
\end{abstract}

\section{Introduction}

\begin{figure}[ht!]
\centering
\includegraphics[width=8.cm]{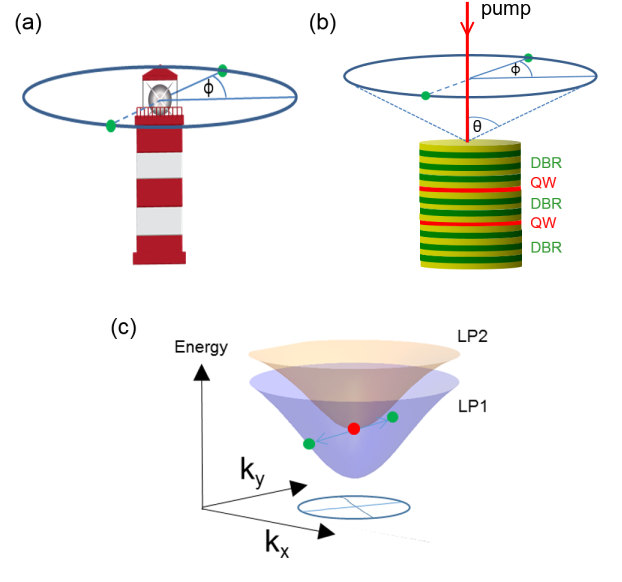}
\caption[graph]{Geometry of the lighthouse emission (a) and of the planar double microcavity (b) composed of three distributed Bragg reflectors (DBR) surrounding two sets of quantum wells (QWs). The microcavity is excited at normal incidence from the top and the emission direction is governed by the incident linear polarization of the resonant excitation. (c) Dispersion of the two lower polariton branches of the planar double microcavity, and representation of the triply degenerate optical parametric oscillation process at work in the lighthouse effect  between LP1 and the LP2 elastic circle. 
}
\label{fig:Fig1}
\end{figure}

Lighthouses have been used for millennia to inform ships on their relative position on the sea. The lighthouse design possesses two advantages (Fig.~\ref{fig:Fig1}a): Its highly directive radiation pattern allows limiting the required power to reach remote locations, and the dynamic control of the emission allows converting spatial information into time information, and vice versa. Such design is used in our everyday life in bar code readers, phased array radars or beamformer sonars.

Non-mechanical lighthouse designs are attractive because they allow reducing mechanical fatigue limitations, and increasing the lifetime, simplicity and speed of devices.
There are two strategies to reach such goal: The first one relies on using the interference of a controllable array of antennas or light modulators. This method is inspired from the phased array radar technology  \cite{visser2006array} and usually requires complex photonic circuits in the visible wavelengths  \cite{sun2013large,jarrahi2008optical,hulme2015fully, hutchison2016high}, controlled electronically by phase shifters. 

Another approach relies on using the nonlinear light-matter response to steer light, opening the road to all-optical operation. Steering can be achieved by using the light field intensity - usually of an auxiliary pump field - to modify the field propagation within the device. Beam steering has been achieved in this way to steer solitons \cite{cao1994optimization, rosberg2006demonstration}, intense fields in biased photorefractive crystals \cite{shwartz2004self}, femtosecond pulses on metallic mirrors \cite{wheeler2012attosecond}, or to control the emission of spatial multimodal lasers by using the nonlinear gain response \cite{liew2014active, bittner2018dynamical}.

The nonlinear approach has the advantage of much simpler and robust designs, it is also usually much faster. But since direction is not controlled independently of intensity, it is hard to imagine practical applications, e.g., in information technology, ranging or microscopy. In this letter we show that, rather surprisingly, the linear polarization is sufficient to achieve continuous control of the beam emission. We demonstrate the lighthouse-like emission of a planar multiple microcavity device (Fig. \ref{fig:Fig1}). This simple device is composed of two coupled microcavities containing multiple quantum wells \cite{ardizzone-etal.13}. The strong coupling between cavity photons and quantum well excitons is reached and the collective excitations of the device are microcavity exciton-polaritons \cite{weisbuch1992observation,sanvitto2012exciton, kavokin2017microcavities}, which are part-light, part-matter quasiparticles and therefore exhibit strong nonlinear interactions. \cite{kuwata1997parametric, savvidis2000angle, huang2000experimental, ciuti2000theory, stevenson2000continuous, langbein2004spontaneous} The dispersion of the two lower polariton branches (LP) is shown in figure \ref{fig:Fig1}c: When we resonantly pump the LP2 at normal incidence, a triply degenerate $\chi^{3}$ parametric scattering process occurs towards LP1, and light is emitted at a finite emission angle $\theta$.

In principle, due to the rotational invariance of the polaritonic dispersion, parametric scattering can occur in any direction, but we predict and observe that in the nonlinear system the emission direction is controlled by the linear pump polarization in the optical parametric oscillation (OPO) regime allowing beam steering over an interval of the azimuth angle $\phi$ of \unit{360}{\degree}.


\section{Theory}
\label{sec:theory}

To describe the scattering of LP2 polaritons at zero in-plane wave vector onto the LP1 elastic circle, we utilize the nonlinear coupled cavity and exciton field theory discussed in Ref.~\cite{kwong-etal.16josab}. 

The core ingredients to this theory are interband polarization functions, obtained from a fermionic quantum-field theory and specialized to excitonic polarizations. The Coulomb interaction between charge carriers gives rise to spin-dependent exciton-exciton interactions and thus to two-exciton correlations, as for example obtained in the dynamics-controlled truncation formalism~\cite{axt-stahl.94,takayama-etal.02}.
The nonlinear optical response of the two subcavities follows from a numerical solution of a coupled mode theory including the excitonic interband polarizations coupled to single-mode equations for the light fields at the positions of the quantum wells, Eqs. (1)-(4) in Ref.~\cite{kwong-etal.16josab}, with
parameter values, in the notation of Ref.~\protect\cite{kwong-etal.16josab}, 
$\gamma_x = \gamma_c = \unit{0.2}{meV}$, $m_{TM} = \unit{0.23}{meV . ps^2 . \micro m^{-2}}$, $m_{TE} = 1.001~m_{TM}$, $\Omega_x = 6.35~\text{meV}$, $\Omega_c =\unit{5.05}{meV}$, $T^{++} = \unit{5.69 \times 10^{-3}}{meV \micro m^2}$, $T^{+-} = -T^{++}/3$, $A_{\mathrm{PSF}} = \unit{2.594 \times 10^{-4}}{\micro m^2}  $; detuning of the cavity mode from the bare exciton frequency $\Delta_P = \unit{-4.3086}{meV}$.

\begin{figure}[ht]
\centering
\includegraphics[width=8.cm]{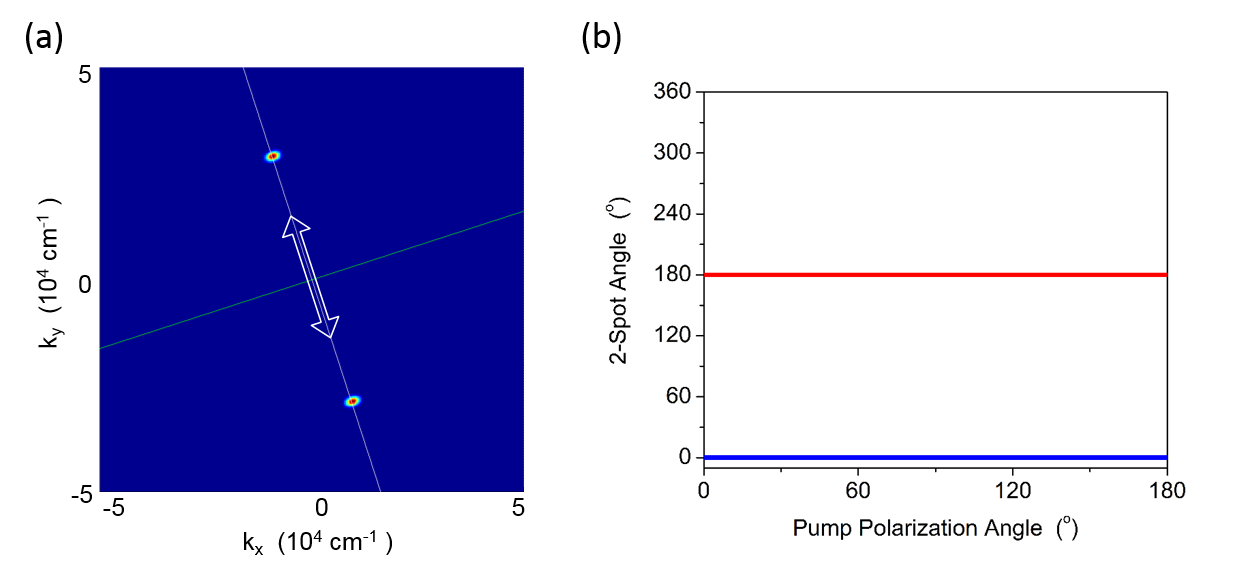}
\caption[graph] {Theoretical simulation of the polaritonic lighthouse effect. 
(a) Two-spot far-field pattern detected in the cross-linear polarization channel for resonant excitation with a linearly polarized cw laser 
(polarization direction in this example is 108$^{\circ}$ relative to the x-axis).
(b) Relative orientation (azimuth) of the two emission intensity maxima with respect to the linear vectorial polarization of the pump. When rotating the pump polarization orientation, the 2-spot emission rotates with it.}
\label{fig:Fig2}
\end{figure}

The nonlinearities entering the theory are phase-space filling and spin-dependent exciton-exciton T-matrices, which are different in the co-circular $T^{++}$ and cross-circular $T^{+-}$ scattering channels. 
The  transverse-electric (TE) and transverse-magnetic (TM) bare cavity modes are assumed to have parabolic dispersion with different curvatures (or cavity effective masses).
This results in polaritonic TE-TM splitting \cite{schumacher-etal.07prb}. 
In the experimental conditions, the two subcavities are slightly different, which leads to a different coupling with the external pump. A quantitative estimate of this difference follows from a linear transfer matrix simulation of the entire structure and as a result the external pump field is chosen to be 
larger (by a factor of $z=-1.52$) in the first subcavity than in the second subcavity. The factor $z$ is defined by the relation
$p^x_{\text{pump},1} = z p^x_{\text{pump},2}$, where 
$p^x_{\text{pump},i}$
are pump parameters for the two subcavities
$i=1,2$. The relation between $p^x_{\text{pump},j}$ and
$R^{\pm}_{\text{pump},i}$ in 
Eqs. (1) of Ref.~\cite{kwong-etal.16josab} is as follows. We first solve the steady-state versions of Eqs. (1) and (2) of Ref.~\cite{kwong-etal.16josab}, Fourier-transform from real (configuration) to wave vector space and take $k = 0$ to describe the spatially homogeneous pump process. These equations read
$ E^{\pm}_{\text{pump},i} = - (
\Delta_p + i \gamma_x 
- T^{++} |p^{\pm}_{\text{pump},i} |^2
- T^{+-} |p^{\mp}_{\text{pump},i} |^2 )
p^{\pm}_{\text{pump},i} /
[  \Omega_x (1 - 2 A_{PSF} |p^{\pm}_{\text{pump},i} |^2 ) ]
$
and
$ R^{\pm}_{\text{pump},i} = - ( \Delta_p + i \gamma_c )
 E^{\pm}_{\text{pump},i}
 + \Omega_c E^{\pm}_{\text{pump},j} 
 + \Omega_x p^{\pm}_{\text{pump},i} 
$.
For given $p^{\pm}_{\text{pump},i}$ they determine $ R^{\pm}_{\text{pump},i} $.
For our linearly polarized pump, we have
$p^{\pm}_{\text{pump},i} = 
\frac{1}{ \sqrt{2}  }
e^{\mp i \phi}  p^{x}_{\text{pump},i}
$
where $\phi$ is the pump polarization angle with respect to the x-axis.
More details of the pump simulation are given 
in Sec. 2.3 of Ref. \cite{luk-thesis.18}.
We use
$p^x_{\text{pump},1} = \unit{12}{\micro m^{-1}}$ 
and $p^x_{\text{pump},2} = p^x_{\text{pump},1}/z$.
For a linearly polarized pump with negative detuning from the bare exciton resonance, because of the spin-dependent exciton-exciton interactions, the polarization channel cross-linear to the pump's polarization is favored for instability (OPO) of pump-induced LP1 polaritons. Together with the TE-TM cavity mode splitting this leads to a spatial anisotropy for polariton scattering. Taking also into account the effect of the cavity difference in the double-cavity system, here we find stable 2-spot patterns at pump powers not too far above the OPO threshold. 
The spatial orientation of these 2-spot patterns is found to be parallel to the vectorial polarization direction of the pump. A representative numerical solution that has reached a steady state for each pump polarization is shown in Fig. \ref{fig:Fig2}. 

The pattern shape depends on the choice in the cavity asymmetry and pump power: The 2-spot pattern of figure~\ref{fig:Fig2} is obtained for a particular choice in the cavity asymmetry ($z \ne 1$) and the pump power. If we choose a symmetric cavity or larger pump powers, we can also obtain a 6-spot pattern with a \unit{60}{\degree} angle between the output fields. Which angles can occur in principle is determined by the nonlinearity, in our case a 3$^{rd}$- order (or $\chi^{(3)}$) nonlinearity. The physical processes giving rise to the  \unit{180}{\degree} (2-spot) and  \unit{60}{\degree} angles (6-spot) are illustrated in Ref. \cite{kwong-etal.17PhysicaScripta}.
For a conventional single-cavity design, we have also found the possibility of \unit{90}{\degree} angles \cite{lewandowski-etal.17optexpress} (4-spot) in the case of linearly polarized pump beams. These \unit{90}{\degree} angles patterns are made possible by the interplay between the exciton-exciton interaction and TE-TM splitting. Interestingly, Whittaker et al. report on the experimental observation of a diverse family of polariton patterns, including ones with an odd number of lobes, by using circularly polarized excitation\cite{whittaker-etal.17}. In their experiment as well, a pattern rotation was observed by changing from left to right circularly polarized pumping. 
We finally note that our model is for an optically isotropic crystal without strain or defects, so that the crystal orientation does not come into play in the selection of patterns and in their orientation.


\section{Experimental setup and observations}
\label{sec:experiment}

\begin{figure}[ht]
\centering
\includegraphics[width=14.cm]{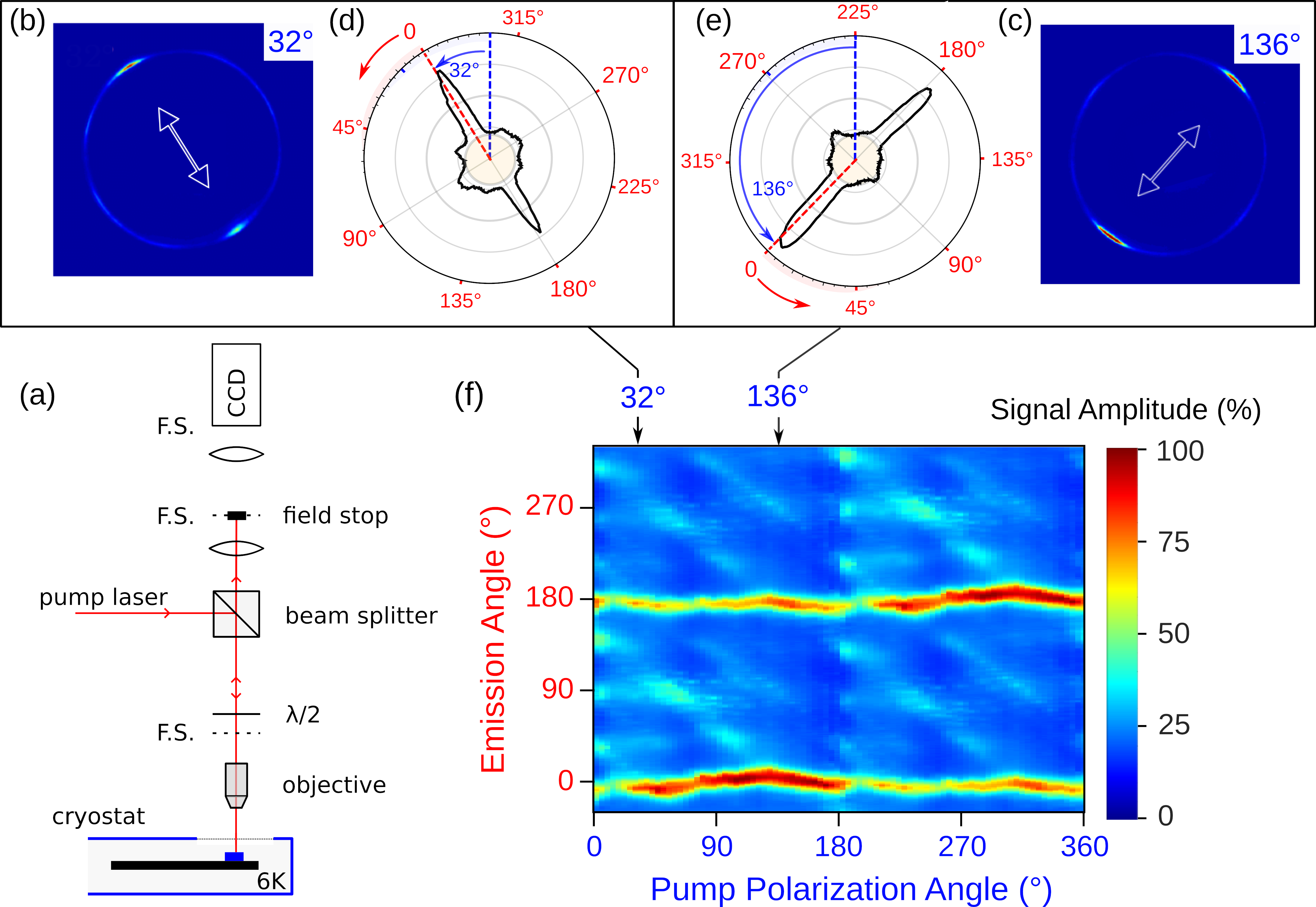}
\caption[graph]{Experimental observation of the polaritonic lighthouse effect. (a) Scheme of the experimental setup: the linear pump polarization is controlled with a half-wave plate, and a field stop is used in the Fourier space (F.S.) to filter the direct pump reflection on the sample. (b),(c) Examples of far field observed for an incoming linear polarization pump beam oriented at \unit{32}{\degree} (resp. \unit{136}{\degree}) with respect to the [100] axis. Two spots are observed collinear to the polarization direction. (d), (e) Corresponding radiation pattern at \unit{32}{\degree} (gray) (resp. \unit{136}{\degree}). Beam width is about \unit{14}{\degree} and extinction reaches up to \unit{10}{dB_i}, the direction of emission is controlled with a \unit{5}{\degree} accuracy. The beige disk represents the direction-independent background due to elastic scattering. 
(f) Map of the emission amplitude as a function of the pump linear polarization angle and of the emission angle relative to the pump polarization. 
}
\label{fig:Fig3}
\end{figure}

The sample, which has been used in  several studies  \cite{ardizzone-etal.13, lafontControllingOpticalSpin2017, lewandowskiPolarizationDependenceNonlinear2016,lukTheoryOpticallyControlled2018}, consists of two coupled $\lambda/2$ Ga$_{0.05}$Al$_{0.95}$As cavities embedded between three Ga$_{0.05}$Al$_{0.95}$As/Ga$_{0.8}$Al$_{0.2}$As distributed Bragg reflectors with 25 (back), 17.5 (middle), and 17.5 (front) pairs respectively (fig. \ref{fig:Fig1}b).
The nominal Q factor is $10^5$, and the middle Bragg mirror induces a \unit{10}{meV} coupling between bare cavity modes. In each cavity, three sets of four \unit{7}{nm} GaAs QWs are inserted at the antinodes of the  field, resulting in a \unit{13}{meV} Rabi splitting.
Experiments are performed at $T=\unit{6}{\kelvin}$, the sample is excited at normal incidence, in resonance with the lower polariton branch 2 (see fig. \ref{fig:Fig1}c), with a linearly polarized single-mode Ti:Sapphire laser. The pump polarization is controlled by a half-wave plate, and the far field is recorded in reflection geometry in the polarization orthogonal to the pump. The pump reflection is spatially filtered.

With a \unit{50}{\micro\meter}-wide resonant excitation beam, we observe the optical parametric oscillation (OPO) regime above an incoming laser power of \unit{60}{mW} and the structuration of the far field in a variety of light patterns  \cite{ardizzone-etal.13}. The two-spot patterns are the simplest of them and are observed just above OPO threshold, when the LP2 branch is slightly negatively detuned with respect to the bare exciton energy  (\unit{-2}{meV}). 

Figure \ref{fig:Fig3} represents the observed far-field emission of the microcavity (\ref{fig:Fig3}a) and the corresponding radiation pattern (\ref{fig:Fig3}b) that is observed at a radial angle $\theta$ of \unit{20}{\degree}. This lighthouse-like emission is always directed along the linear pump polarization, independently of the crystal orientation, as evidenced by the direction emission map in Fig. \ref{fig:Fig3}c.


\section{Discussion}
\label{sec:discussion}

Let us now discuss the experimental observations in light of the theoretical prediction of a lighthouse-like emission.  Since the polariton lighthouse effect implies both a directional and controllable emission, we are going to analyze these two aspects.

The radiation patterns are approximately symmetric and highly directional: In the radial direction, the lighthouse emission is well focused with a beam width $\Delta \theta$ of only \unit{0.4}{\degree}. This feature results from the strong constraint on radial emission set by the phase-matching conditions on resonant parametric scattering between the two lower polaritonic branches. In the azimuthal direction, the beam width is $\Delta \phi = 14^{\circ}$ independent of the excitation linear polarization configuration. 

In contrast, the theoretical prediction on figure \ref{fig:Fig2} suggests much tighter azimuthal focusing is possible. This discrepancy probably originates from the effect of elastic scattering on line defects within the DBRs, which are signaled by a characteristic speckle signature in the elastically scattered signal (not shown). Indeed, whereas polariton-polariton parametric scattering in a uniform planar microcavity preserves total momentum, additional random elastic scattering lifts this restriction.
 
Most of the emission is cast in the two main lobes. The main lobes apparent imbalance results from a partial field stop within the large NA collecting objective. Stray light  accounts for \unit{[-7;-10]}{dB_i} and has two origins: First, the resonant elastic scattering on DBR defects, which forms a direction independent background (yellow disk on fig. \ref{fig:Fig3}b)  \cite{lewandowskiPolarizationDependenceNonlinear2016}. Second, the competing parametric scattering pathways which are responsible for the formation of structured side lobes. 
Figure \ref{fig:Fig3}b) gives a representative example of radiation pattern (incoming polarization at \unit{32}{\degree} with the [100] crystalline axis), as well as the most selective configuration (\unit{136}{\degree}). From figure \ref{fig:Fig3}c), we see that side lobes are most pronounced when the incoming polarization is oriented along the [100] direction. This anisotropic response is most probably the result of residual built-in strain effects within the heterostructure.\cite{dasbach2002oscillations, lafont2016origins}
We note that a clear lighthouse effect is evidenced over 90\% of the incoming polarization angles and that the two-spot emission always dominate side lobes emission.

The angular precision of the lighthouse emission by the linear polarization is achieved with a remarkably high accuracy of \unit{5}{\degree}. 
Previous work using InGaAs/GaAs microcavities presented strong anisotropic line defects and mosaicity which pinned the OPO emission on the semiconductor crystalline axes\cite{abbarchiDiscretizedDisorderPlanar2012}. Therefore, we conclude that the small lattice mismatch of AlGaAs/GaAs microcavities is critical to achieve controllability of the emission.

The maximum emitted power in the main pattern reaches \unit{200}{\micro W}, which is only $0.4\%$ of the incoming power (\unit{60}{mW}, necessary to reach OPO threshold). This currently low external power efficiency (emitted power/incident power) is due to the fact that a strong Kerr effect occurs at resonance. As a result, the microcavity LP2 resonance is repelled from the excitation laser energy and most of the incoming power gets reflected on the first DBR and does not enter the microcavity. \cite{baas2004optical} 
By taking into account the low injection efficiency, we estimate that the internal power efficiency (emitted power/transmitted power) is in the $25\%$ range, whereas the theoretical limit is $50\%$ due to power balance between parametrically scattered beams in the OPO regime.


\section{Conclusion}
\label{sec:conclusion}

We have shown that an optical microcavity can present lighthouse-like emission continuously controlled by the linear vectorial polarization of the pump laser itself. The use of vectorial polarization as a control parameter implies that this function can only be implemented in a nonlinear optical device. 

We used a coupled cavity theory combined with a many-particle theory describing exciton-exciton interactions to demonstrate the polariton lighthouse effect if a minimal set of physical ingredients are included, namely the nonlinear optical response of microcavity exciton-polaritons, the difference in the subcavities forming the microcavity, and the rotational invariance of the device. 
We observed the lighthouse effect with a device made of quantum wells embedded in a double microcavity and evidenced that a two-spot pattern oriented along the linear incoming pump polarization is observed for most azimuth angles with good control on the emission direction.

With our present cavity, we observe some deviations from the ideal lighthouse emission, such as parasitic stray light, originating from elastic scattering on line defects and the effect of built-in strain, and inherent to real world devices. Progresses in heterostructures quality and power efficiency could turn this polariton lighthouse effect into an original and useful all-optical beam-steering method which can be of great interest for microscopy or LIDARs.

\begin{acknowledgement}

We acknowledge that this project has received funding from the Agence Nationale de la Recherche (ANR) (ANR-16-CE24-0023 TeraMicroCav), the US National Science Foundation (NSF) (DMR 1839570), and the Deutsche Forschungsgemeinschaft (SCHU 1980/5-2 and Heisenberg grant 270619725). RB acknowledges CPU time at HPC (University of Arizona); SS acknowledges computing time at Paderborn Center for Parallel Computing ($\mathrm{PC^2}$).

\end{acknowledgement}

\begin{suppinfo}

The microcavity geometry is detailed in section 3. During fabrication by molecular beam epitaxy, a wedge in the cavity thicknesses is introduced by stopping the cavity rotation during growth of the cavity layers. As a consequence, the sample has a natural gradient in the bare cavity energy, while the confined exciton energy is mostly unaffected. Note that upper and lower cavities have a parallel gradient in energy so that the coupled cavity modes inherit the same energy gradient. By recording the photoluminescence of lower polariton branches 1 and 2 on the sample surface, we can reconstruct the polariton anticrossing curve and characterize the polariton key properties of the sample. Figure \ref{fig:FigSupp}a shows a typical dispersion curve obtained by photoluminescence, where only the two lower polariton branches (labeled LPB1 and LPB2) are visible. Lower polariton energies at normal incidence are represented function of the position on the sample along the gradient direction on figure \ref{fig:FigSupp}b, the position corresponding to panel a is indicated as green points.

\begin{figure}[ht]
\centering
\includegraphics[width=14.cm]{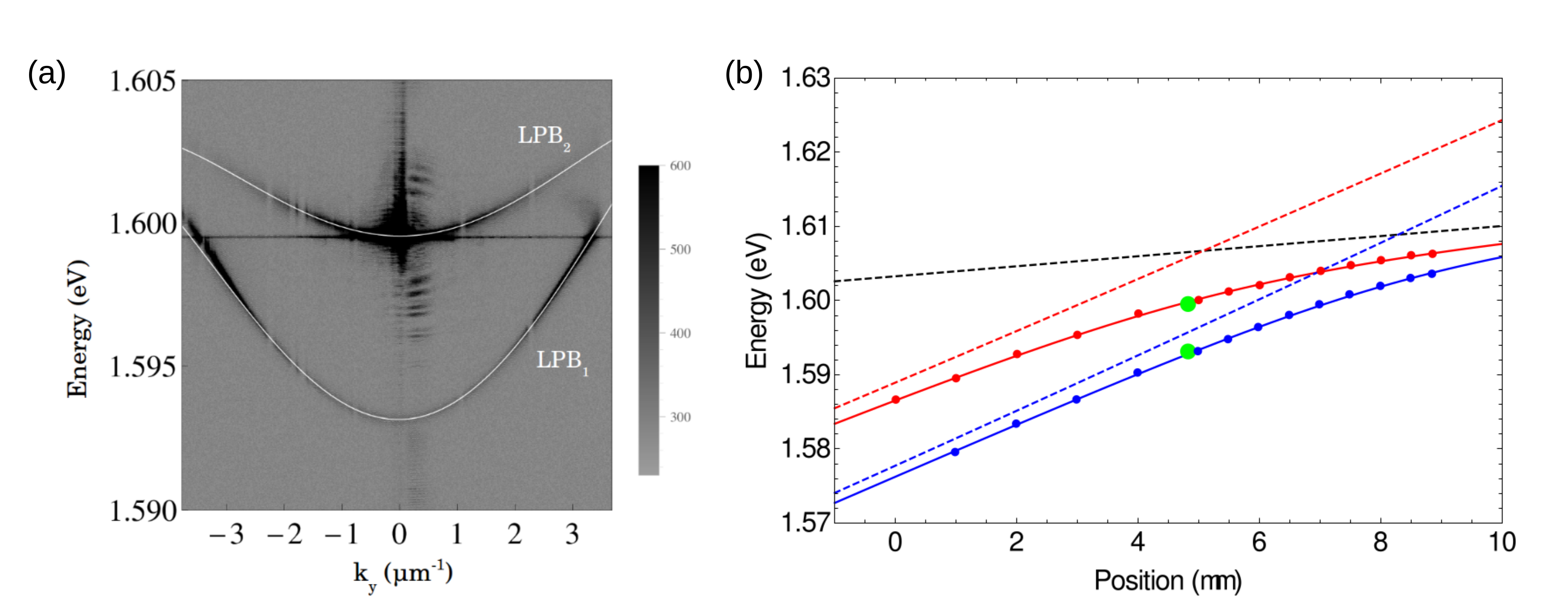}
\caption[graph]{Characterization of the double microcavity exciton-polaritons (a) Typical energy dispersion for the double microcavity sample. Only the two lower polaritons branches are visible (LPB1 and LPB2). The pump is resonant with LPB2 at normal incidence. The white lines correspond to a
fourth order polynomial fit of the branches to obtain as precisely as possible the
energy minima of LPB1 and LPB2. (b) Anticrossing curve obtained from photoluminescence spectra 
at various positions onto the sample. The two green points correspond to the position
of panel (a). The dashed black line corresponds to the bare exciton energy as a function of the position on the sample.The blue
and red dashed lines correspond to the optical cavity modes’ energies as a function of the position on the sample.
}
\label{fig:FigSupp}
\end{figure}

The two lower polariton branches of figure \ref{fig:FigSupp}b are fitted by 
$$ E_\mathrm{pol}^{(i)} = \frac{1}{2}(E_{\mathrm{cav}}^{(i)}+E_{\mathrm{exc}}+\sqrt{(E_{\mathrm{cav}}^{(i)}-E_{\mathrm{exc}})^2+4\Omega_R^2}) $$
where $E_{\mathrm{cav}}^{(i)}$ is the coupled cavity mode forming the polariton of interest, and both $E_{\mathrm{cav}}$ and $E_{\mathrm{exc}}$ evolves linearly with position on the sample. 
From these data we infer the cavity energy gradient of \unit{3.6 \pm 0.1}{meV.mm ^ {-1}}, the exciton energy $\unit{1.606\pm 0.003}{eV}$, and the Rabi coupling $\Omega_R=\unit{6.4 \pm 0.8}{meV}$, which is half the Rabi splitting between the lower and upper polaritons. From the coupled cavity splitting, we also deduce the coupling between the upper and lower microcavities $\Omega_c = \unit{5.1\pm 0.7}{meV}$. 

\end{suppinfo}

\bibliography{lighthouse}




%
%
%

\end{document}